\documentstyle[12pt]{article}
\newcommand{\be}{\begin{equation}}
\newcommand{\ee}{\end{equation}}

\newcommand{\al}{\alpha}

\newcommand{\g}{\gamma}
\newcommand{\Om}{\Omega}
\newcommand{\om}{\omega}

\newcommand{\G}{\Gamma}

\newcommand{\La}{\Lambda}

\newcommand{\la}{\lambda}

\newcommand{\dz}{\wedge}

\newcommand{\ba}{\begin{array}}
\newcommand{\ea}{\end{array}}
\newcommand{\beq}{\begin{eqnarray}}
\newcommand{\eeq}{\end{eqnarray}}
\newtheorem{lm}{Lemma}
\newtheorem{th}{Theorem}
\newtheorem{pr}{Proposition}
\newtheorem{co}{Corollary}
\newtheorem{rem}{Remark}
\newtheorem{deff}{Definition}
\newcommand{\bd}{\begin{deff}}
\newcommand{\ed}{\end{deff}}
\newcommand{\bl}{\begin{lm}}
\newcommand{\el}{\end{lm}}
\newcommand{\bp}{\begin{pr}}
\newcommand{\ep}{\end{pr}}
\newcommand{\bt}{\begin{th}}
\newcommand{\et}{\end{th}}
\newcommand{\bc}{\begin{co}}
\newcommand{\ec}{\end{co}}
\newcommand{\brm}{\begin{rem}}
\newcommand{\erm}{\end{rem}}

\newcommand{\der}{\mbox{d}}
\begin{document}

\thispagestyle{empty}

\title {ON A CERTAIN FORMULATION OF THE EINSTEIN EQUATIONS
\footnote{Research supported in part by: Komitet Bada\'n Naukowych 
(Grant nr 2 P03B 017 12), Consorzio per lo Sviluppo Internazionale 
dell'Universita degli Studi di Trieste and 
Erwin Schr\"{o}dinger 
International Institute for Mathematical Physics.}\\ 
\vskip 1.truecm
{\small {\sc Pawe\l  ~Nurowski}}\\
{\small {\it Dipartimento di Scienze Matematiche }}\\
\vskip -0.3truecm
{\small {\it Universita degli Studi di Trieste, Trieste, Italy}}
\footnote{Permanent address: {\it Katedra Metod Matematycznych Fizyki, 
Wydzia\l ~Fizyki, Uniwersytet Warszawski, ul. Ho\.za 74, Warszawa, 
Poland, e-mail: nurowski@fuw.edu.pl}}}
\author{\mbox{}}
\maketitle
\begin{abstract}
We define a certain differential system on an open set of ${\bf R}^6$.
The system locally defines a Lorentzian 4-manifold satisfying the Einstein 
equations. The converse statement is indicated and its details are 
postponed to the furthcoming paper. 
\end{abstract}
\newpage
\noindent
Let $\cal U$ be an open subset of ${\bf R}^6$. Suppose that on 
$\cal U$ we have six 1-forms $(\Lambda, F, \bar{F}, $ $T,E,\bar{E})$ which 
satisfy the following conditions 
\begin{itemize}
\item[{\bf i)}]
$\La,T$ are real- and $F,E$ are complex-valued 1-forms
\item[{\bf ii)}] 
$\La\dz F\dz\bar{F}\dz T\dz E\dz\bar{E}\neq 0$ at each point $p$ of $\cal U$
\item[{\bf iii)}] there exist complex-valued 1-forms $\Om$ and $\G$ on 
$\cal U$, and a certain complex function $\al$ on $\cal U$ such that 
$$
{\rm d}F=(\Om -\bar{\Om})\dz F+E\dz T+\G\dz\La
$$
\be
{\rm d}T=\bar{\G}\dz F + \G\dz\bar{F}-(\Om +\bar{\Om})\dz T\label{eq:sys1}
\ee
$$
{\rm d}\La=\bar{E}\dz F+ E\dz\bar{F}+(\Omega+\bar{\Omega})\dz\La
$$
$$
{\rm d}E=2\Om\dz E+\bar{F}\dz T+\al \La\dz F.
$$
\end{itemize}
The aim of this paper is to prove that we can associate to such $\cal U$ 
a Lorentzian 4-manifold that satsifies the Einstein equations.
The proof will be a direct consequence of the following sequence of 
(very simple) Lemmas.
 
\bl~\\
If $(\Lambda, F, \bar{F},T,E,\bar{E})$ satisfy (\ref{eq:sys1}) then there 
exist complex functions $a,h$ on $\cal U$ and a real constant $\la$ such that  
$$
{\rm d}\G=2\G\dz\bar{\Om}+\bar{\al}T\dz F+\bar{a}(T\dz\La
-F\dz\bar{F})+h\La\dz\bar{F}
$$
\be
{\rm d}\Om=E\dz\bar{\G}-(\al+\frac{\la}{2})(T\dz\La+F\dz\bar{F})+a\La\dz F.
\label{eq:sys2}
\ee
\el
The proof of this Lemma is straightforward but lenghty. It is performed by 
imposing conditions 
\be
\der^2\La=\der^2 T=\der^2F=\der^2 E=0
\label{eq:dru1}
\ee 
on the system (\ref{eq:sys1}) and by using the independence condition 
{\bf ii)}. Application of (\ref{eq:dru1}) implies the possible forms of 
$\der\G$ and $\der\Om$. Use of 
\be
\der^2\G=\der^2\Om=0
\label{eq:dru2}
\ee 
shows, in particular, that $\la$ must be a real constant. $\Box$\\ 
There are also other 
equations that are implied by (\ref{eq:dru1})-(\ref{eq:dru2}). They carry  
information about the differentials of $\al, a, h$ and about the 
decompositions of $\G$ and $\Om$ onto the basis 1-forms 
$(\Lambda, F, \bar{F},T,E,\bar{E})$. Explicitely we have 
\beq
&\der\al=-2aE-\bar{\g}_3T+\bar{\g}_1\bar{F}+sF-w\La,\nonumber\\
&\der a=\bar{h}E-(3\al+\la)\bar{\G}-2a\Om +sT-w\bar{F}+rF-t\La,
\label{eq:domy1}\\
&\der h=-4h\bar{\Om}+4\bar{a}\G-\bar{r}T+\bar{t}F-u\La+v\bar{F},\nonumber
\eeq
where the possible forms of $\Om$ and $\G$ are 
\beq
&\Om=-\om_1\La+\om_2F+\om_3\bar{F}+\om_4T,\label{eq:domy2}\\
&\G=\g_1\La+4\bar{\om}_1F-\g_3\bar{F}-
4\bar{\om}_2T-(3\bar{\al}+\la)\bar{E}.\nonumber
\eeq
Here $\g_1, \g_3,s,w,r,t,u,v,\om_1,\om_2,\om_3,\om_4$ are certain 
complex functions on $\cal U$.\\
Note that the differential system (\ref{eq:sys1}), (\ref{eq:sys2}), 
(\ref{eq:domy1}), (\ref{eq:domy2}) is still not closed. For example, 
the equations of the sort $\der^2\al=0$ should be still imposed.\\

\bl~\\
$\cal U$ is locally foliated by 2-dimensional manifolds ${\cal S}_x$, which 
are tangent to the real distribution $\cal V$ defined by 
$$
\La ({\cal V})=F({\cal V})=T({\cal V})=0.
$$
\el
Here the proof is an immediate application of the Froebenius theorem, since 
the forms $(\La, F, \bar{F}, T)$ form a closed differential ideal due to the 
relations (\ref{eq:sys1}). $\Box$\\
Let us define 
\be
G=F\bar{F}-T\La 
\ee
on $\cal U$. $G$ constitutes a degenerate metric on $\cal U$. It has the 
signature (+,+,+,--,0,0).
\bl~\\
$G$ is constant along any leaf ${\cal S}_x$ of the 
foliation $\{{\cal S}_x\}$.
\el  
To proof this we define a basis of vector fields 
$(Y,f,\bar{f},X,e,\bar{e})$, which is the respective dual of  
$(\Lambda, F, \bar{F},T,E,\bar{E})$. Then we notice that the distribution 
$\cal V$ is spanned by vector fields of the form 
$$
V=Ue+\bar{U}\bar{e},
$$
where $U$ is any complex function on $\cal U$. Using an arbitary $V$ and 
the explicit form of $G$ one easily finds that ${\cal L}_V G=0$ due to the  
equations (\ref{eq:sys1}) and the properties of the Lie derivative  
${\cal L}_V$. $\Box$\\
Now, we introduce an equivalence relation $\sim$ on $\cal U$ which 
identifies points lying on the same leave of $\{{\cal S}_x\}$. We assume 
that the quotient space ${\cal M}={\cal U}/\sim$ is a 4-manifold. 
According to Lemma 3, $G$ projects down to a well defined nondegenerate 
Lorentzian metric $g$ on $\cal M$. 
\bt~\\
The Lorentzian metric $g$ on ${\cal M}={\cal U}/\sim$ satisfies the Einstein 
equations $R_{ij}=\la g_{ij}$ and is not conformally flat. 
\et    
Proof.\\
To proof the theorem, we first consider a  
4-dimensional submanifold ${\cal M}'$ of $\cal U$ that is transversal to the 
leaves of $\{{\cal S}_x\}$. We have a natural inclusion 
$\iota :{\cal M}'\hookrightarrow\cal U$. ${\cal M}'$ may be equipped with a 
Lorentzian metric $g'=\iota^*G$. It is clear that $({\cal M}, g)$ and 
$({\cal M}',g')$ are locally isometrically equivalent. Thus, we 
may represent $({\cal M},g)$ by $({\cal M}',g')$. In this way, it is enough 
to show that $g'$ satisfies the Einstein equations. To do this, we first 
observe that $g'=\iota^*(F)\iota^*(\bar{F})-\iota^*(T)\iota^*(\La)$. Thus, 
a set of forms 
$\theta^i=(\iota^*(F),\iota^*(\bar{F}),\iota^*(T),\iota^*(\La))$, 
(i=1,2,3,4) constitutes a null cotetrad for $g'$. To calculate the 
curvature we need to know $\der\theta^i$. But these, due to the fact that 
$\iota^*\der =\der\iota^*$, are given by the realtions (\ref{eq:sys1}). 
To simplify the notation, we will omit the 
signs of the pullback $\iota^*$ in all of the formulae. On doing that we 
find that the connection 1-forms for $g'$ determined by the equations
$$
\der F=-\G^1_{~1}\dz F-\G^1_{~3}\dz T-\G^1_{~4}\dz \La
$$
$$
\der \bar{F}=-\G^2_{~2}\dz\bar{F}-\G^2_{~3}\dz T-\G^2_{~4}\dz \La 
$$
$$
\der T=-\G^3_{~1}\dz F-\G^3_{~2}\dz\bar{F}-\G^3_{~3}\dz T 
$$
$$
\der\La=-\G^4_{~1}\dz F-\G^4_{~2}\dz\bar{F}-\G^4_{~4}\dz \La 
$$
$$
g_{ik}\G^{k~}_j+g_{jk}\G{k~}_i=0, \quad\quad
g_{ij} =\left(\begin{array}{cccc}
0&1&0&0\\
1&0&0&0\\
0&0&0&-1\\
0&0&-1&0
\end{array}\right)
$$
read
$$
\G^i_{~j} =\left(\begin{array}{cccc}
\bar{\Om}-\Om&0&-E&-\G\\
0&\Om-\bar{\Om}&-\bar{E}&-\bar{\G}\\
-\bar{\G}&-\G&\Om+\bar{\Om}&0\\
-\bar{E}&-E&0&-\Om-\bar{\Om}
\end{array}\right).
$$
The curvature ${\cal R}^i_{~j}=\der\G^i_{~j}+\G^i_{~k}\dz\G^k_{~j}$ 
of this connection can be calculated 
using  the relations (\ref{eq:sys1}), (\ref{eq:sys2}). Modulo the obvious 
reality conditions its components read as follows.
\beq
&{\cal R}^1_{~3}=\al F\dz\La-\bar{F}\dz T\nonumber\\
&{\cal R}^1_{~4}=-\bar{\al}T\dz F+\bar{a}(\La\dz T+
F\dz\bar{F})+h\bar{F}\dz\La\nonumber\\
&\frac{1}{2}({\cal R}^3_{3}+{\cal R}^1_{1})=
(\bar{\al}+\frac{1}{2}\la)(\La\dz T +F\dz \bar{F})+\bar{a}\La\dz\bar{F}.
\nonumber
\eeq
These should be compared with the definitions of the spinorial coeficients 
of the Weyl tensor $\Psi_\mu$, $\mu=0,1,2,3,4$, the traceless Ricci tensor  
$S_{ij}$ and the Ricci scalar $R$ given by 
\beq                 
&{\cal R}^1_{~3}=\bar{\Psi}_4 \bar{F}\dz T+\bar{\Psi}_3 
(\La\dz T-F\dz \bar{F})+
(\bar{\Psi}_2 +\frac{1}{12}R)\La\dz F\nonumber\\
&+\frac{1}{2}S_{33}F\dz T+\frac{1}{2}S_{32}(\La\dz T+F\dz\bar{F})+
\frac{1}{2}S_{22}\La\dz\bar{F},\nonumber\\
&\quad\nonumber\\
&{\cal R}^1_{~4}=(-\Psi_2-\frac{1}{12}R)F\dz T-
\Psi_1(\La\dz T+F\dz \bar{F})-
\Psi_0 \La\dz \bar{F}\nonumber\\
&-\frac{1}{2}S_{22}\bar{F}\dz T-\frac{1}{2}S_{42}(\La\dz T-F\dz\bar{F})-
\frac{1}{2}S_{44}\La\dz F,\nonumber\\
&\quad\nonumber\\
&\frac{1}{2}({\cal R}^3_{~3}+{~\cal R}^1_{1})=-\Psi_3 F\dz T+
(-\Psi_2+\frac{1}{24}R) 
(\La\dz T+F\dz \bar{F})-\Psi_1 \La\dz \bar{F}\nonumber\\
&-\frac{1}{2}S_{32}\bar{F}\dz T-\frac{1}{4}(S_{12}+S_{34})
(\La\dz T-F\dz\bar{F})-\frac{1}{2}S_{41}\La\dz F.\nonumber
\eeq~\\
Looking at these equations we easily get  
$S_{ij}=R_{ij}-\frac{1}{4}Rg_{ij}\equiv 0$. This proves 
the Einstein property of the metric.\\
The spinorial coeficients for the Weyl tensor are also 
easy to obtain. They read   
$$
\Psi_0=h,~~~\Psi_1=-\bar{a},~~~\Psi_2=-\bar{\al} -\frac{\la}{3},~~~
\Psi_3=0,~~~\Psi_4=-1.
$$
Due to the nonvanishing of $\Psi_4$ the metric is never conformally flat.\\
We can also interpret the constant $\la$ which for the 
first time appeared in equations 
(\ref{eq:sys2}). One easily reads from the above that it is proportional 
to the Ricci scalar $R=4\la$.\\
This concludes the proof of the Theorem. $\Box$\\
We, additionally, note that the metric is of the Cartan-Petrov-Penrose 
type D iff 
\be
a=0, ~~~~~~{\rm and}~~~~~ h=-(3\bar{\al}+\la)^2 \label{eq:D}
\ee
and of the type N iff
\be
h=a=0,~~~~~~{\rm and}~~~~~\la=-3\bar{\al}.\label{eq:N}
\ee
The metric is algebraically special iff $I^3=27J^2$, where 
$$
I=-h+\frac{1}{3}(3\bar{\al}+\la)^2, ~~~{\rm and}~~~J=\bar{a}^2+
\frac{1}{27}(3\bar{\al}+\la)^3+\frac{1}{3}h(3\bar{\al}+\la).
$$
The converse of the present paper can be stated in the following theorem 
(see Ref. \cite{bi:nur} for details).
\bt~\\
Let $\cal M$ be a 4-dimensional manifold with a Lorentzian metric $g$ 
satisfying the Einstein equations  $R_{ij}=\la g_{ij}$. Then, there exists a 
double branched cover $\tilde{\cal P}$ of the bundle of null directions 
${\cal P}$ over $\cal M$ {\rm \cite{bi:nur1,bi:nur2}}
\begin{itemize}
\item[i)] which is a fibration over $\cal M$ with fibers being 
2-dimensional tori (or, in algebraically special cases, their 
degenerate counterparts {\rm \cite{bi:nur}}) 
\end{itemize}
and 
\begin{itemize}
\item[ii)] on which there exists a unique system of 1-forms 
$(\Lambda, F, \bar{F}, T,E,\bar{E})$ which satisfies the following three 
conditions 
\begin{itemize}
\item[{\bf i)}]
$\La,T$ are real- and $F,E$ are complex-valued 1-forms
\item[{\bf ii)}]
$\La\dz F\dz\bar{F}\dz T\dz E\dz\bar{E}\neq 0$ at each point $p$ of 
$\tilde{\cal P}$
\item[{\bf iii)}]
$$
{\rm d}F=(\Om -\bar{\Om})\dz F+E\dz T+\G\dz\La
$$
$$
{\rm d}T=\bar{\G}\dz F + \G\dz\bar{F}-(\Om +\bar{\Om})\dz T
$$
$$
{\rm d}\La=\bar{E}\dz F+ E\dz\bar{F}+(\Omega+\bar{\Omega})\dz\La
$$
$$
{\rm d}E=2\Om\dz E+\bar{F}\dz T+\al \La\dz F.
$$
with certain complex-valued 1-forms $\Om$ and $\G$, and a certain 
complex function $\al$ on $\tilde{\cal P}$. 
\end{itemize} 
\end{itemize}
\et

\noindent
{\Large{\bf Acknowledgements}}\\ 
I warmly thank Andrzej Trautman for stimulating my 
interest in the papers \cite{bi:Cart1,bi:Cart2,bi:Pen}.\\ 
I am also very grateful to Mike Crampin, Simonetta Fritelli, 
Jerzy Lewandowski, Lionel Mason, Ted Newman, David Robinson and 
Helmuth Urbantke for helpful discussions.

\end{document}